\documentclass{article}
\usepackage{graphicx}
\usepackage{color}
\def\beq{\begin{equation}}
\def\eeq{\end{equation}}
\usepackage{graphicx}
\usepackage{amsmath}
\usepackage{amsfonts}
\usepackage{amssymb}
\usepackage{amscd}

\begin{document}
\begin{center}
{\bf Euclidean relativistic quantum mechanics - 
scattering asymptotic conditions%
}
\end{center}

\bigskip

\begin{center}
{
W. N. Polyzou \\
The University of Iowa \\
Gordon Aiello\\
The University of Iowa 
}
\end{center}

\bigskip

\noindent{\bf abstract:}
We discuss the formulation of the scattering asymptotic 
condition in a relativistic quantum theory formulated in terms
of reflection positive Euclidean Green functions.

\bigskip

\noindent{\bf Introduction}

\bigskip

Schwinger \cite{Schwinger:pna} showed that the spectral condition implies
that the Green functions of a local quantum field theory have an analytic
continuation to Euclidean times.  This observation has motivated
Euclidean formulations of quantum field theory that have been used
successfully in numerical lattice discretizations of the theory.  One
difficulty with the Euclidean formulation is that the imaginary time
leads to problems in formulating scattering problems
\cite{Maiani} that involve asymptotic time limits.  Osterwalder and
Schrader \cite{Osterwalder:1973dx} proved a reconstruction theorem that showed how to
construct a relativistic quantum field theory from a set of Euclidean
invariant Green functions satisfying a condition called reflection
positivity.  The proof of the reconstruction theorem has three
important attributes.  First, it is constructive, leading to an
explicit representation of the physical Hilbert space and a set of
self-adjoint operators on that space satisfying the Poincar\'e Lie
algebra.  Second, the locality property is logically independent of
the other properties that are needed to construct a relativistic
quantum theory.  This implies that it should be possible to make
approximations or truncations of a local theory that still retain all
of the remaining properties of a relativistically invariant quantum
theory.  Third, the construction of the relativistic quantum theory
does not require explicit analytic continuation of the Euclidean
times. This paper is based on \cite{Aiello:2015jgc}\cite{polyzou:2015}
where we argue that given a
Hilbert space representation and a set of Poincar\'e generators on that
space that satisfy cluster properties, it should be possible to
formulate scattering directly in the Euclidean representation of the
Hilbert space.

\bigskip
\noindent{\bf Reconstruction of quantum mechanics}
\bigskip

A dense set of vectors in the Hilbert space consists of sets of
normalizable functions of Euclidean space-time variables,
\begin{equation}
\mathbf{f}= (f_0, f_1(x_{e11}), f_2 (x_{e21},x_{e22}),\cdots ), 
\end{equation}
with support for positive relative Euclidean times, 
$0 < x^0_{en1} < x^0_{en2} < x^0_{en3} \cdots$.

The Hilbert-space inner product of two such vectors 
is expressed in terms of a quadratic form involving the
Euclidean Green functions, $G^e_{n}(x_{e1}, \cdots ,x_{en})$, by
\begin{equation}
\langle \mathbf{f} \vert \mathbf{g} \rangle =
\end{equation}
\begin{equation}
\sum_{mn} \int f_m^* ( x_{em1}, 
\cdots ,  x_{emm})
G^e_{m+n}( T_ex_{emm}, \cdots , T_e x_{em1}, y_{en1}, \cdots ,y_{enn})
\times
\label{a1}
\end{equation}
\begin{equation}
g_n (y_{en1}, \cdots , y_{enn}) d^{4m}x d^{4n}y 
\end{equation}
where $T_e$ Euclidean time reflection.  The collection of Euclidean 
Green functions are called reflection positive if   
$\langle \mathbf{f} \vert \mathbf{f} \rangle \geq 0$ for all 
$\mathbf{f}$ satisfying the support condition.  This represents 
the physical Hilbert space inner product, even though all of the integrals
in (\ref{a1}) are over Euclidean variables and there is no analytic 
continuation.
This inner product has zero norm vectors, so Hilbert space vectors are
actually equivalence class of functions $\mathbf{f}$ subject to the 
equivalence relation $\mathbf{f} \sim \mathbf{g}$ if and only if
$\langle \mathbf{f}-\mathbf{g} \vert \mathbf{f}- \mathbf{g} \rangle =0$.

The infinitesimal generators of the Poincar\'e group are represented 
by the following self-adjoint operators on this space
\begin{equation}
H f_n (x_{en1}, x_{en2},\cdots , x_{enn})
=
\sum_{k=1}^n {\partial \over \partial x_{ekn}^0} f_n 
(x_{en1}, x_{en2},\cdots , x_{enn})
\label{a2}
\end{equation}
\begin{equation}
\mathbf{P}  f_n (x_{en1},x_{en1},\cdots , x_{1nn})
=
-i \sum_{k=1}^n {\partial \over \partial \mathbf{x}_{enk}} 
f_n (x_{en1}, x_{en2},\cdots , x_{enn})
\label{a3}
\end{equation}
\begin{equation}
\mathbf{J} f_n ( x_{en1}, x_{en2},\cdots ,
x_{ne}) = -i \sum_{k=1}^n \mathbf{x}_{enk} \times {\partial \over
\partial \mathbf{x}_{enk}} f_n (x_{en1},
x_{en2},\cdots ,x_{enn})
\label{a4}
\end{equation}
\[
\mathbf{K} f_n (x_{en1}, x_{en2},\cdots ,
x_{ne}) = 
\]
\begin{equation}
\sum_{k=1}^n (\mathbf{x}_{enk} \times {\partial \over
\partial {x}^0_{enk}}-  {x}^0_{enk}  {\partial \over
\partial \mathbf{x}_{enk}}) f_n ( x_{en1},
x_{en2},\cdots , x_{enn}).
\label{a5}
\end{equation}
It is straightforward to show that these operators are Hermetian with respect
to the inner product (\ref{a1}) and satisfy the Poincar\'e commutation
relations.  They are also self-adjoint by virtue of being generators
of unitary one-parameter groups, contractive semigroups, or local
symmetric semigroups \cite{Klein:1981}\cite{Klein:1983}\cite{Frohlich:1983kp} 
on this Hilbert space.  
The Hamiltonian $H$ also satisfies a spectral condition as a consequence 
of reflection positivity \cite{glimm:1981}.

If the Euclidean Green functions satisfy cluster properties,
\begin{equation}
\lim_{\mathbf{a} \to \infty} 
(G^e_{m+n} (X_{em}+\mathbf{a},Y_{en},\cdots) - G^e_m (X_{em}) G^e_n (Y_{em})) \to 0 
\end{equation}
\begin{equation}
X_{em} = (x_{em1},\cdots ,x_{emm})
\end{equation}
then the Poincar\'e generators defined by (\ref{a2}-\ref{a5}) also 
satisfy cluster properties.

\bigskip
\noindent{\bf Formulation of scattering}
\bigskip

Scattering matrix elements,
$S_{fi} = \langle \mathbf{f}_+ \vert \mathbf{f}_- \rangle$, are
defined as the inner product of initial and final states at a given
time that asymptotically look like systems of non-interacting
particles long before and long after a collision.  The relation
between these states and the corresponding asymptotic states is given
by the scattering asymptotic condition, which replaces the initial
condition on the initial or final states by the scattering asymptotic
conditions
\begin{equation}
\lim_{t \to \pm \infty} 
\Vert e^{-iHt} \vert \mathbf{f}_{\pm} \rangle -
\Phi e^{-iH_{f} t} \vert \chi_{} \rangle \Vert =0 .
\label{a6}
\end{equation}
The quantity $\vert \chi_{}\rangle$ on the right describes 
free-particle wave packets.  For a two-particle final state it has the form
\begin{equation}
\Phi e^{-iH_{f12} t} \vert \chi_{12} \rangle := 
\sum_{\mu_1\mu_2} \int \underbrace{\vert \phi_{b1}, \mathbf{p}_1, \mu_1, 
\phi_{b2}, \mathbf{p}_2, \mu_2 \rangle}_{\Phi} \times
\end{equation}
\begin{equation}
\underbrace{e^{-i (E_1 (\mathbf{p}_1) + E_2(\mathbf{p}_2))t}}_{e^{-iH_{f12}t}}
d\mathbf{p}_1 d\mathbf{p}_2
\underbrace{\chi_1(\mathbf{p}_1, \mu_1)\chi_2(\mathbf{p}_1, \mu_1)}_{wave \, packets}
\end{equation}
where
\begin{equation}
M_{bi} \vert \phi_{bi} \rangle = m_{b_i} \vert \phi_{bi} \rangle \qquad
E_i(\mathbf{p}_i )= \sqrt{m_{b_i}^2 + \mathbf{p}_i^2 }
\end{equation}
and $m_{b_i}$ is in the point-spectrum of the subsystem mass operator.
It follows that the operator $\Phi$ is a mapping from a two-particle
asymptotic Hilbert space of square integrable functions
of $\mathbf{p}_1,\mu_1,\mathbf{p}_2.\mu_2$, 
to the physical Hilbert space with inner product (\ref{a1}).

A sufficient condition for the existence of these limits is the Cook 
condition, which in this notation has the form
\begin{equation}
\int_a^\infty
\Vert (H\Phi- \Phi H_{f}) e^{\mp iH_{f} t} \vert \chi \rangle \Vert
dt < \infty
\end{equation}
where $(H\Phi- \Phi H_{f})$ replaces the short-range potential in the 
original formulation of Cook's method\cite{Cook:1957}.

This condition can be formulated in the Euclidean representation.
Two conditions are needed to establish convergence in the Euclidean
case.  First, the Green functions need to satisfy cluster properties.
Second, the operator $\Phi$ must eliminate the contribution from the
disconnected parts of the Green function in (\ref{a6}).

To illustrate how this works consider the case of $2-2$ scattering.  Cluster 
properties lead to a four-point Green function that can be
expressed as the sum of products of two-point functions and a 
connected fourpoint function:
\begin{equation}
G_4 = G_{2e}G_{2e}+G_{4ec}.
\end{equation} 
In general there will be additional disconnected terms - the analysis below
can be easily extended to treat these cases.

In the Euclidean representation the square of integrand of (\ref{a6}) for the 
case of 2-2 scattering has the form  
\[
\Vert (H \Phi - \Phi H_f ) 
e^{-i H_{f} t} \vert \chi_{0\pm} (0) \rangle \Vert^2 =
\]
\[
(\chi_f ,e^{i H_f t} (\Phi^{\dagger}H-H_f\Phi^{\dagger}),   
T_e (G_{2e}G_{2e}+G_{4ec}) (H\Phi-\Phi H_f) e^{-i H_i t}\chi_i)  =
\]
%
\[
\int \chi_1^* (\mathbf{p}_1) \chi_2^* (\mathbf{p}_2) 
e^{i  (E_{m_1}(\mathbf{p}_1) + E_{m_2}(\mathbf{p}_2))t}
d\mathbf{p}_1
d\mathbf{p}_2 \times
\]
\[
(
{\partial \over \partial x^{0}_1} + {\partial \over \partial x^{0}_2}
- 
E_{m_1}(\mathbf{p}_1) - E_{m_2}(\mathbf{p}_2))
\langle \mathbf{p}_1 , \mathbf{p}_2  \vert \Phi^{\dagger} 
\vert x_1 , x_2 \rangle
\times 
\]
\[
d^4x_1 d^4 x_2 
(G_{2e} (T_e x_1, y_1)
G_{2e} (T_e x_2, y_2)+
G_{4ec} (T_e x_1, T_e x_2, y_2, y_1))
d^4 y_1 d^4 y_2 \times
\]
\[
(
{\partial \over \partial y^{0}_1} + {\partial \over \partial y^{0}_2}
- 
E_{m_1}(\mathbf{p}_1') - E_{m_2}(\mathbf{p}_2'))
\langle y_1 , y_2 \vert \Phi  \vert \mathbf{p}_1' , \mathbf{p}_2'  \rangle
\times
\]
\begin{equation}
e^{-i  (E_{m_1}(\mathbf{p}_1') + E_{m_2}(\mathbf{p}_2'))t}
\chi_1 (\mathbf{p}_1') \chi_2 (\mathbf{p}_2') d\mathbf{p}_1'
d\mathbf{p}_2'
\label{main}
\end{equation}
where $E_{m_i}(\mathbf{p})= \sqrt{m_i^2 +\mathbf{p}^2}$ are the energies
of the outgoing particles, and $G_{4ec}$ is the connected part of 
the Euclidean 4-point Green function.

We expect that the contribution to this expression from the connected
Green functions will fall off like $1/t^3$ for large time, as it does
in the Minkowski case (see \cite{jost} p. 132).
For the disconnected terms it is necessary to
study the contribution from two-point Green functions.  These can be
expressed in terms of the Kall\"en-Lehmann representation of the
Euclidean two-point function:
\[
\langle \xi \vert \psi \rangle =  
\int \xi^*(x_e) G_{2e}(T_e x_e-y_e) \psi (y_e) d^4x_e d^4y_e =
\]
\[
\int {d^4 p_e d^4x_e d^4y_e \over (2 \pi)^4}
\xi ^*(x_e)
{ e^{i p_e^0 (-x_e^0-y_e^0) +i \mathbf{p} \cdot (\mathbf{x}-\mathbf{y})}
\rho (m)
\over (p_e^0)^2 + \mathbf{p}^2 + m^2 }  \psi(y_e)  dm 
=
\]
\begin{equation}
\int
\underbrace{\xi^*(x_e) {d^4x_e \over (2 \pi)^{3/2}}
e^{- E_m(\mathbf{p})x_e^0 +i \mathbf{p} \cdot \mathbf{x}}
}_{\tilde{\xi}^* (\mathbf{p},m)} 
{\rho (m) dm d\mathbf{p}
\over 2 E_m(\mathbf{p})  } 
\underbrace{
e^{- E_m(\mathbf{p})y_e^0 -i \mathbf{p} \cdot \mathbf{y}}
{ d^4y_e\over (2 \pi)^{3/2}} \psi(y_e)
}_{\tilde{\psi}(\mathbf{p},m)} .
\label{ip}
\end{equation}
The Euclidean time support condition of the wave functions
$\psi(y_e)$ and $\xi^*(x_e)$
allows the $p_0$
integral to be computed by contour integration.

The subsystem mass operator can be seen to be the Euclidean 4-Laplacian
by integration by parts in (\ref{ip}): 
\begin{equation}
\int 
{\rho (m) 
\over 2 E_m(\mathbf{p})}   
e^{- E_m(\mathbf{p})y_e^0 -i \mathbf{p} \cdot \mathbf{y}}
{ d^4y_e\over (2 \pi)^{3/2}} \nabla^2_{ye}\psi(y_e)=   
\end{equation}
\begin{equation}
\int 
{\rho (m)
\over 2 E_m(\mathbf{p})}   
e^{- E_m(\mathbf{p})y_e^0 -i \mathbf{p} \cdot \mathbf{y}}
{ d^4y_e\over (2 \pi)^{3/2}} m^2\psi(y_e)   
\end{equation}
The problem is to find subsystem point mass eigenstates
$\psi(y_e)$ and $\xi(x_e)$ with the mass of the asymptotic particle
that are also consistent with the Euclidean time-support condition.

The spectrum of the subsystem mass operator is identical to the support
of the Lehmann weight, $\rho(m)$.  The Lehmann weight is assumed to  
have the form
\beq
\rho(m)= \sum z_i \delta (m-m_i) + \rho_{ac}(m)
\label{del}
\eeq
where the support of the continuous part, $\rho_{ac}(m)$ of $\rho(m)$
is a half line that starts at the sum of the masses of the lightest
possible intermediate particles. 

In order to select the desired asymptotic
particle it is necessary to choose wave functions $\psi (x_e)$ that only
get contributions from the Lehmann weight corresponding to mass of the
asymptotic particle.  This corresponds to one of the delta functions 
in (\ref{del}).  

One way to do this, motivated by the method used by Haag and Ruelle
\cite{Haag:1958vt}\cite{Ruelle:1962}\cite{jost}, 
is to multiply the wave function by a function
$h(m^2)$ that is 1 when $m^2$ is the square of the mass of the asymptotic
particle and 0 on the rest of the Lehmann weight, $\rho(m^2)$.  Since
the square of the cluster mass operator is the Euclidean Laplacian,
a wave function of the form 
\begin{equation}
\psi (x_e) = h(\nabla^2)  g(x_e)
\end{equation}
has the desired properties provided it satisfies the support condition.
Here  $g(x_e)$ is another function satisfying the support condition.
The simplest choice of $g(x_e)$ satisfying the support condition is
a function of the form
\begin{equation}
g(x_e) = \delta (x^0_e- \tau) \tilde{g}(\mathbf{x}).
\end{equation}
It is interesting that in spite of presence of the delta function that
enforces the support condition, this is a normalizable vector
with respect to the scalar product (\ref{a1}).

The question is whether the support conditions on $h(m^2)$ and
$h(\nabla^2) g(x_e)$ are compatible.  The concern is that on one hand
an infinite series of derivatives, like a translation operator, can
change support conditions. On the other hand $h(m^2)$ cannot even be
be analytic if the Lehmann weight has any continuous spectrum because
$h(m^2)$ it must vanish on the continuous support of the Lehmann
weight.

In the unphysical situation, where the support of the Lehmann weight
consists of a finite number of discrete points, a finite-degree
polynomial $h(\nabla^2)$ can be constructed that selects the desired
mass and automatically preserves the support condition.  Also, in the
case where the support of the continuous portion of the Lehmann weight
is compact, the Weierstrass approximation theorem implies that
$h(\nabla^2)$ can be uniformly approximated by a polynomial.  While both
cases are unphysical, both preserve the support condition of the wave
function.

This leads to the question of whether it is possible to make a
polynomial approximation to $h(m^2)$ that converges to any
continuous function of the half line.  It turns out that there
is a sufficient condition called  the Carleman condition that
provides a suitable sufficient (but not necessary) condition.

The concern is that applying $h(\nabla^2)$ to $g(x_e)$ does not
automatically preserve the positive Euclidean-time support
condition of $g(x_e)$. 

The Carleman condition \cite{Carleman:1926}\cite{Akheizer:1965}gives a sufficient condition
\begin{equation}
\sum_{n=1}^\infty \gamma_n^{-{1 \over 2n}} > \infty  
\end{equation}
on an
infinite collection of moments $\gamma_n$ on the half line
\begin{equation}
\gamma_n = \int_0^{\infty} w(x)dx x^n
\end{equation}
to uniquely solve for the measure, $w(x)dx$.

When this condition is satisfied, multiplication by $x$ has a
unique-self adjoint extension on the Hilbert space with
weight $w(x)$ and the orthogonal polynmials with
respect to the weight $w(x)$ are complete.
 
For the Euclidean two-point function the relevant moments are
\begin{equation}
\gamma_n := ( \psi , T_e G_{2e}  
(\nabla^2)^n \psi )  =
\int_0^\infty w(m) m^{2n} .
\end{equation}
For the scattering application we actually need the completeness of
polynomials in $m^2$ rather than $m$ which is why the $n^{th}$
moment involves $m^{2n}$.

For the two-point example above the relevant moments are
\begin{equation}
\gamma_n := \int_0^\infty {e^{- \sqrt{m^2 + \mathbf{p}^2} \tau} 
\over 2 \sqrt{m^2 + \mathbf{p}^2}} \rho(m)
m^{2n} dm .
\end{equation}
In order to estimate these moments we assume 
(see theorem 6.2.4 in \cite{glimm:1981}) that the
continuous part of the Lehmann weight is polynomially bounded.
In this case it is sufficient to replace the moments by
\begin{equation}
\gamma_n \to \gamma_n' = 
\int_0^\infty {e^{- \sqrt{m^2 + \mathbf{p}^2} \tau} 
\over 2 \sqrt{m^2 + \mathbf{p}^2}}  
m^{2n+k} dm .
\end{equation}
where $k$ is a constant (the polynomial bound).
With some algebra it is possible to prove the inequality
\begin{equation}
\sum_{n=1}^\infty \gamma_n^{\prime -{1 \over 2 n}} > 
\sum_{n=0}^\infty  {c \over 2n +k-2} > \infty
\end{equation}
which is sufficient to show that $h(\nabla^2)$ can be approximated by
a polynomial, thus ensuring that it is possible to construct 
point-spectrum mass eigenstates that satisfy the Euclidean 
time support condition.

Returning to equation (\ref{main}) it is now possible to see
how the condition above eliminates the disconnected terms in this
equation.  The first step is to replace the wave packets
$\chi(x_e)$ by $h_n(\nabla^2)g(x_e)$, where $h_n$ is 
a polynomial approximation to $h(m^2)$.  Integrating the 
Laplacians by parts means that
they can be replaced by $h_n(m^2)$, where $m^2$ is integrated
over the support of the Lehmann weight.  
The completeness of the
polynomials in $m^2$ means that we can make the contribution
from the continuous part of the Lehmann weight as small as desired.
Integrating the Euclidean time derivatives by parts in the
two-point functions leads to replacements of the form
${\partial \over \partial x^0} \to \omega_m (\mathbf{p})$.
When this is integrated over $m$ the only term that is
picked up is the $m=m_i$ selected by $h(m^2)$.  This term cancels the
energy factors that come from $\Phi H_f$.  What remains is just the
contribution from the connected four-point Green function, which
is expected to vanish at $t$ gets large.

This shows that it is possible to compute $S$-matrix elements
in the Euclidean representation.  As a practical matter $e^{iHt}$ is not
easy to compute in the Euclidean representation.  There are a number
of approaches that can be used given a representation for
$H$.  One that we have tested makes use of the fact that given a set of 
reflection positive Green functions, matrix elements of
polynomials in $(e^{-\beta H})$
are easy to calculate using the inner product (\ref{a1}).   

To take advantage of this note that the invariance principle
\cite{kato:1966}\cite{Chandler:1976} implies
\beq
\lim_{t \to \pm \infty} e^{iHt} \Phi e^{-iH_ft} \vert \chi \rangle
=
\lim_{n \to \pm \infty} e^{-in e^{-\beta H}} \Phi e^{in e^{-\beta H_f}} \vert \chi \rangle .
\eeq
Because the spectrum of $e^{-\beta H}$ is compact, for any fixed $n$,
$e^{-in e^{-\beta H}}$ can be uniformly approximated by a polynomial
in $e^{-\beta H}$.  Matrix elements of $e^{-\beta m H}$ simply
translate the argument of the Euclidean wave functions by $-\beta m$.
In principle $\beta$ is any positive number; in practice it sets the
energy scale of the problem.  We have demonstrated, in a solvable model,
that these methods can be applied to calculate sharp momentum cross
sections over a wide range of energies \cite{Kopp:2011vv}.  There are
many other possibilities given that the reconstruction theorem
provides all of the elements
necessary to formulate the scattering problem.

This work upported by the US DOE office of science - award number DE-SC0016457.


\begin{thebibliography}{27}
%
%

\expandafter\ifx\csname natexlab\endcsname\relax\def\natexlab#1{#1}\fi
\expandafter\ifx\csname bibnamefont\endcsname\relax
  \def\bibnamefont#1{#1}\fi
\expandafter\ifx\csname bibfnamefont\endcsname\relax
  \def\bibfnamefont#1{#1}\fi
\expandafter\ifx\csname citenamefont\endcsname\relax
  \def\citenamefont#1{#1}\fi
\expandafter\ifx\csname url\endcsname\relax
  \def\url#1{\texttt{#1}}\fi
\expandafter\ifx\csname urlprefix\endcsname\relax\def\urlprefix{URL }\fi
\providecommand{\bibinfo}[2]{#2}
\providecommand{\eprint}[2][]{\url{#2}}

\bibitem{Schwinger:pna}
\bibinfo{author}{\bibfnamefont{J.~S.} \bibnamefont{Schwinger}},
  \bibinfo{journal}{Proc. Natl. Acad. Sci. U. S.}
  \textbf{\bibinfo{volume}{44}}, \bibinfo{pages}{956} (\bibinfo{year}{1958}).

\bibitem{Maiani}
\bibinfo{author}{\bibfnamefont{L.}~\bibnamefont{Maiani}} \bibnamefont{and}
  \bibinfo{author}{\bibfnamefont{M.}~\bibnamefont{Testa}},
  \bibinfo{journal}{Phys. Lett. B} \textbf{\bibinfo{volume}{245}},
  \bibinfo{pages}{585} (\bibinfo{year}{1990}).

\bibitem{Osterwalder:1973dx}
\bibinfo{author}{\bibfnamefont{K.}~\bibnamefont{Osterwalder}} \bibnamefont{and}
  \bibinfo{author}{\bibfnamefont{R.}~\bibnamefont{Schrader}},
  \bibinfo{journal}{Commun. Math. Phys.} \textbf{\bibinfo{volume}{31}},
  \bibinfo{pages}{83} (\bibinfo{year}{1973}).

\bibitem{Aiello:2015jgc}
\bibinfo{author}{\bibfnamefont{Gordon}~\bibnamefont{Aiello}} \bibnamefont{and}
  \bibinfo{author}{\bibfnamefont{W.~N.}~\bibnamefont{Polyzou}},
  \bibinfo{journal}{Phys.Rev.} \textbf{\bibinfo{volume}{D93}},
  \bibinfo{pages}{056003} (\bibinfo{year}{2016}), \eprint{1512.03651}.

\bibitem{polyzou:2015}
\bibinfo{author}{\bibfnamefont{W.~N.}~\bibnamefont{Polyzou}},
  \bibinfo{journal}{Phys.Rev.} \textbf{\bibinfo{volume}{D89}},
  \bibinfo{pages}{076008} (\bibinfo{year}{2014}), \eprint{1312.03585}.
  
\bibitem{Klein:1981}
\bibinfo{author}{\bibfnamefont{A.}~\bibnamefont{Klein}} \bibnamefont{and}
  \bibinfo{author}{\bibfnamefont{L.}~\bibnamefont{Landau}}, \bibinfo{journal}{J.
  Functional Anal.} \textbf{\bibinfo{volume}{44}}, \bibinfo{pages}{121}
  (\bibinfo{year}{1981}).

\bibitem{Klein:1983}
\bibinfo{author}{\bibfnamefont{A.}~\bibnamefont{Klein}} \bibnamefont{and}
  \bibinfo{author}{\bibfnamefont{L.}~\bibnamefont{Landau}}, \bibinfo{journal}{Comm.
  Math. Phys} \textbf{\bibinfo{volume}{87}}, \bibinfo{pages}{469}
  (\bibinfo{year}{1983}).

\bibitem{Frohlich:1983kp}
\bibinfo{author}{\bibfnamefont{J.}~\bibnamefont{Frohlich}},
  \bibinfo{author}{\bibfnamefont{K.}~\bibnamefont{Osterwalder}},
  \bibnamefont{and} \bibinfo{author}{\bibfnamefont{E.}~\bibnamefont{Seiler}},
  \bibinfo{journal}{Annals Math.} \textbf{\bibinfo{volume}{118}},
  \bibinfo{pages}{461} (\bibinfo{year}{1983}).

\bibitem{glimm:1981}
\bibinfo{author}{\bibfnamefont{J.}~\bibnamefont{Glimm}} \bibnamefont{and}
  \bibinfo{author}{\bibfnamefont{A.}~\bibnamefont{Jaffe}},
  \emph{\bibinfo{title}{Quantum Physics; A functional Integral Point of View}}
  (\bibinfo{publisher}{Springer-Verlag}, \bibinfo{year}{1981}).

\bibitem{Cook:1957}
\bibinfo{author}{\bibfnamefont{J.}~\bibnamefont{Cook}}, \bibinfo{journal}{J.
  Math. Phys.} \textbf{\bibinfo{volume}{36}}, \bibinfo{pages}{82}
  (\bibinfo{year}{1957}).

\bibitem{jost}
\bibinfo{author}{\bibfnamefont{R.}~\bibnamefont{Jost}},
  \emph{\bibinfo{title}{The General Theory of Quantized Fields}}
  (\bibinfo{publisher}{AMS}, \bibinfo{year}{1965}).

\bibitem{Haag:1958vt}
\bibinfo{author}{\bibfnamefont{R.}~\bibnamefont{Haag}}, \bibinfo{journal}{Phys.
  Rev.} \textbf{\bibinfo{volume}{112}}, \bibinfo{pages}{669}
  (\bibinfo{year}{1958}).

\bibitem{Ruelle:1962}
\bibinfo{author}{\bibfnamefont{D.}~\bibnamefont{Ruelle}},
  \bibinfo{journal}{Helv. Phys. Acta.} \textbf{\bibinfo{volume}{35}},
  \bibinfo{pages}{147} (\bibinfo{year}{1962}).

\bibitem{Carleman:1926}
\bibinfo{author}{\bibfnamefont{T.}~\bibnamefont{Carleman}},
  \emph{\bibinfo{title}{Les fonctions quasi analytiques, Collection de
  Monographies sur la Th\'eorie des Fonctions}}
  (\bibinfo{publisher}{Gauthier--Villars, Paris}, \bibinfo{year}{1926}).

\bibitem{Akheizer:1965}
\bibinfo{author}{\bibfnamefont{N.~I.} \bibnamefont{Akheizer}},
  \emph{\bibinfo{title}{The Classical Moment Problem and some related questions
  in analysis}} (\bibinfo{publisher}{Oliver and Boyd, Edinburgh and London},
  \bibinfo{year}{1965}).


\bibitem{kato:1966}
\bibinfo{author}{\bibfnamefont{T.}~\bibnamefont{Kato}},
  \emph{\bibinfo{title}{Perturbation theory for linear operators}}
  (\bibinfo{publisher}{Spinger-Verlag, Berlin}, \bibinfo{year}{1966}).

\bibitem{Chandler:1976}
\bibinfo{author}{\bibfnamefont{C.}~\bibnamefont{Chandler}} \bibnamefont{and}
  \bibinfo{author}{\bibfnamefont{A.}~\bibnamefont{Gibson}},
  \bibinfo{journal}{Indiana Journal of Mathematics.}
  \textbf{\bibinfo{volume}{25}}, \bibinfo{pages}{443} (\bibinfo{year}{1976}).

\bibitem{Kopp:2011vv}
\bibinfo{author}{\bibfnamefont{P.}~\bibnamefont{Kopp}} \bibnamefont{and}
  \bibinfo{author}{\bibfnamefont{W.~N.}~\bibnamefont{Polyzou}},
  \bibinfo{journal}{Phys.Rev.} \textbf{\bibinfo{volume}{D85}},
  \bibinfo{pages}{016004} (\bibinfo{year}{2012}), \eprint{1106.4086}.

  
  
\end{thebibliography}
\end{document}